\documentclass[%
 aip,
 amsmath,amssymb,
 reprint 
]{revtex4-1}
\usepackage{graphicx}
\usepackage{dcolumn}
\usepackage{bm}

\usepackage[T1]{fontenc}
\usepackage{mathptmx}
\usepackage{etoolbox}
\usepackage{color}
\usepackage[dvipsnames]{xcolor}
\usepackage{stackengine}
\usepackage{subfigure}
\usepackage{siunitx} 
\usepackage[nameinlink,capitalize]{cleveref}
\usepackage{placeins}

\makeatletter
\def\@email#1#2{%
 \endgroup
 \patchcmd{\titleblock@produce}
  {\frontmatter@RRAPformat}
  {\frontmatter@RRAPformat{\produce@RRAP{*#1\href{mailto:#2}{#2}}}\frontmatter@RRAPformat}
  {}{}
}%
\makeatother
\begin{document}

\preprint{}

\title
{Thermodynamic modeling for numerical simulations based on the generalized cubic equation of state }

\author{T. Trummler}
 \affiliation{
  Institute of Applied Mathematics and Scientific Computing, Bundeswehr University Munich\\ 
   Werner-Heisenberg-Weg 39, 85577 Neubiberg, Germany 
}
\author{M. Glatzle}
 \affiliation{
 BooleWorks GmbH, Radlkoferstrasse 2, 81373 Munich, Germany 
}
\author{A. Doehring}
 \affiliation{
  Institute of Applied Mathematics and Scientific Computing, Bundeswehr University Munich\\ 
   Werner-Heisenberg-Weg 39, 85577 Neubiberg, Germany 
}
\author{N. Urban}
 \affiliation{
  Institute of Applied Mathematics and Scientific Computing, Bundeswehr University Munich\\ 
   Werner-Heisenberg-Weg 39, 85577 Neubiberg, Germany 
}
\author{M. Klein}
 \affiliation{
  Institute of Applied Mathematics and Scientific Computing, Bundeswehr University Munich\\ 
   Werner-Heisenberg-Weg 39, 85577 Neubiberg, Germany 
}

\email{theresa.trummler@unibw.de}

\date{\today}

\begin{abstract}
  
  We further elaborate on the generalized formulation for cubic equation of state proposed by Cismondi and Mollerup [Fluid Phase Equilib. 232 (2005) 74--89]~\citep{Cismondi}. With this formulation all well-known cubic equations of state can be described with a certain pair of values, which allows for a generic implementation of different equations of state. Based on this generalized formulation, we derive a complete thermodynamic model for computational fluid dynamics (CFD) simulations by providing the resulting correlations for all required thermodynamic properties. For the transport properties, we employ the Chung correlations. 

  Our generic implementation includes the often used equations of state Soave--Redlich--Kwong and Peng--Robinson and the Redlich--Kwong--Peng--Robinson (RKPR) equation of state. The first two assume a universal compressibility factor and are therefore only suitable for fluids with a matching critical compressibility. The Redlich--Kwong--Peng--Robinson overcomes this limitation by considering the equation of state parameter as function of the critical compressibility. We compare the resulting thermodynamic modeling for the three equations of state for selected fluids with each other and CoolProp reference data.  

  As supplementary material to this paper, we provide a Python tool called real gas thermodynamic python library (\texttt{realtpl}). This tool can be used to evaluate and compare the results for a wide range of different fluids. Additionally, we also provide the implementation of the generalized form in OpenFOAM. 

\end{abstract}

\maketitle

\section{Introduction}
  Numerical flow simulations of super- and transcritical conditions require appropriate thermodynamic models. A state-of-the-art thermodynamic model is based on a cubic equation of state (EoS) and departure function formalism for the evaluation of enthalpy and energy. This can be found in early~\citep{zong2004numerical,kim2011numerical,park2012and,petit2013large} as well as recent computational fluid dynamics (CFD) investigations~\citep{fathi2022large,traxinger2018experimental,Matheis:2017dx,muller2016large,sharan2021investigation,lagarza2019large,poblador2022volume}. For the cubic EoS mostly the Peng-Robinson~\citep{PR} (PR) or the Soave-Redlich-Kwong EoS~\citep{SRK,soave1972equilibrium} (SRK) are employed. Both assume a universal critical compressibility and are therefore only suitable for fluids with a matching critical compressibility. Volume translation methods~\citep{abudour2012volume,martin1979cubic,harstad1997efficient,matheis2016volume} represent one possible solution to improve the density prediction for fluids, that are not well described by SRK or PR. 
  \citet{Cismondi} suggested the Redlich-Kwong-Peng-Robinson EoS (RKPR), introducing a third EoS parameter and formulating all three EoS parameters as a function of the critical compressibility. Building upon their work, \citet{Kim} presented a thermodynamic modeling approach based on the RKPR and demonstrated its advantages and suitability for different fluids. Despite its advantages, only a few recent studies~\citep{fathi2022large,jung2020real} have employed the RKPR EoS recently for real gas CFD simulations of n-dodecane injections. 
  Within the suggestion of the RKPR, \citet{Cismondi} and earlier Mollerup (see \citet{michelsen2004thermodynamic}), also proposed a general formulation of the cubic EoS by which all of the well-known cubic EoS can be described with a particular set of values. Such a formulation allows for a modularized implementation of all these cubic EoS, thus, less code duplication and a better readability. 

  An alternative to a cubic EoS is the PC-SAFT EoS (perturbed-chain statistical associating fluid theory)~\citep{gross2001perturbed}, which has been successfully employed by \citet{rodriguez2019simulation,rodriguez2018simulation} for two-dimensional CFD simulations. Another approach is the usage of tabulated reference data, yielding higher accuracy~\citep{Koukouvinis} and also a potentially faster evaluation of the thermodynamic data~\citep{Doehring,jafari2022exploring}. However, cubic EoS are still mostly used due to their simplicity and overall good accuracy. In addition to the EoS and the relations for thermodynamic properties, CFD simulations also require relations for the transport properties viscosity and thermal conductivity. \citet{Chung} proposed correlations for these transport properties, which are often employed for such real gas simulations~\citep{matheis2016volume,Matheis:2017dx, Matheis, fathi2022large,traxinger2018experimental, muller2016large}. Alternative methods are for example, the Lucas method~\citep{Poling} for the viscosity and the Stiel--Thodos method~\citep{Poling} for the thermal conductivity. These methods have been recently employed by \citet{sharan2021investigation} for nitrogen and showed a good agreement with NIST reference data. Alternatively, the residual entropy scaling technique can be used for the calculation of the thermal conductivity~\citep{hopp2017thermal} and the viscosity~\citep{looetgering2015group} as done by \citet{Koukouvinis}. In general, it is an important step in CFD simulations to check the accuracy and suitability of a thermodynamic model in advance, as included in several studies~\citep{Matheis:2017dx,Koukouvinis,traxinger2018experimental,doehring2022nicfd}. 
  For different fluids, pressure and temperature ranges, such an evaluation can be complicated and time-consuming. Apart from that, such a verification requires an already successful implementation of the thermodynamic model. This is usually not the case during the development or further development of a CFD solver. The open source library CoolProp~\citep{bell2014pure} provides implementations for the SRK and PR EoS. 
  Bell et al.~\citep{Bell2016Thermo} wrote a comprehensive thermodynamic library to evaluate chemical properties specifically targeted for chemical engineering. Our new tool \texttt{realtpl}, on the other hand, has been specifically designed for applications in the context of CFD simulations and evaluates the entire thermodynamic model required for these simulations. 


  In this paper, we aim to further promote the idea of the generalized formulation of cubic EoS by Mollerup~\citep{Cismondi,michelsen2004thermodynamic}, i.e., one formulation for all three cubic EoS (PR, SRK, and RKPR). To this end, we describe in detail how this formulation is solved and present the resulting relations for the thermodynamic properties. We also outline the overall thermodynamic model based on this generalized formulation. For the thermodynamic model, we employ the Chung correlations for the evaluation of the transport properties. We apply the thermodynamic model to selected fluids and study its suitability. Therewith, we also demonstrate the good applicability of the RKPR for all fluids with different critical compressibility factors. In order to test the proposed thermodynamic model and to apply it to different configurations, we provide an open source Python tool called \texttt{realtpl} for real gas thermodynamic python library. This tool can be used to evaluate and compare the results for a wide range of different fluids. Additionally, we also provide the implementation of the generalized form in OpenFOAM. 

  The paper is structured as follows: \Cref{s:ThermoMod} presents the thermodynamic model based on the generalized cubic EoS. Then, in \cref{s:Accuracy}, the applicability and suitability of the thermodynamic model is assessed for selected fluids comparing the model using SRK, PR, and RKPR. \Cref{s:SupMat} contains information about the additionally provided Python tool \texttt{realtpl} and the validation of the proposed OpenFOAM implementation based on the generalized formulation. Finally, the paper is summarized in \cref{s:Con}. 

\begin{table*}[!tb]
   \caption{Parameters of the EoS adopted from \citet{Kim}.}
  \label{tab:KimTabelle}
  \centering
  \begin{tabular}{p{0.5cm}p{3.7cm}p{3.7cm}p{6.5cm}}
    \hline
      & \textbf{\small SRK \citep{SRK}} & \textbf{\small PR \citep{PR}} & \textbf{\small RKPR \citep{Cismondi}} \\
\hline
    $\delta_{1}$ & $1$     & $1+\sqrt{2}$ & $d_{2}+d_{1}(d_{3}-c_z Z_{c})^{d_{4}}+d_{5}(d_{3}-c_z Z_{c})^{d_{6}}\newline
    \mathrm{with}\; c_z = 1.168, \newline
     d_1 = 0.428363,\; d_2 = 18.496215, \newline
     d_3 = 0.338426,\; d_4 = 0.660000, \newline
      d_5 = 789.723105,\; d_6 = 2.512392$\\
\hline
    $\delta_{2}$ & $0$     & $1-\sqrt{2}$ & $(1-\delta_{1})/(1+\delta_{1})$\\
\hline
    $a$     & $0.42747\left(\frac{{R}^{2}T_{c}^{2}}{p_{c}}\right)$ & $0.45724\left(\frac{{R}^{2}T_{c}^{2}}{p_{c}}\right)$ 
    & $\frac{3y^{2}+3yd+d^{2}+d-1}{(3y+d-1)^{2}}\left(\frac{{R}^{2}T_{c}^{2}}{p_{c}}\right)$ \\
\hline
    $b$     & $0.08664\left(\frac{{R}T_{c}}{p_{c}}\right)$ & $0.0778\left(\frac{{R}T_{c}}{p_{c}}\right)$ & $\frac{1}{3y+d-1}\left(\frac{{R}T_{c}}{p_{c}}\right)$ \\
    \multicolumn{1}{l}{} & \multicolumn{1}{c}{} & \multicolumn{1}{c}{} & \multicolumn{1}{l}{with \enspace $d=\frac{1+\delta_{1}^{2}}{1+\delta_{1}}$} \\
    \multicolumn{1}{l}{} & \multicolumn{1}{c}{} & \multicolumn{1}{c}{} & \multicolumn{1}{c}{$y = 1+[2(1+\delta_{1})]^{\frac{1}{3}}+\left(\frac{4}{1+\delta_{1}}\right)^{\frac{1}{3}}$} \\   
\hline
    $\alpha$ & $\left(1+\kappa(1-\sqrt{T/T_{c}})\right)^{2}$  
             & $\left(1+\kappa(1-\sqrt{T/T_{c}})\right)^{2}$   
             & $(3/(2+T/T_{c}))^{\kappa}$ \\
\hline
    $\kappa$ 
    & $0.48508 \newline + 1.55171\,\omega \newline- 0.15613\,\omega^2$  
    & $0.37464 \newline + 1.54226\,\omega \newline - 0.26992\,\omega^2$ 
    & $(66.125\, c_z Z_{c} - 23.359)\omega^{2}  \newline
       +(-40.594\, c_z Z_{c} + 16.855)\omega   \newline
       +(5.27345\, c_z Z_{c} - 0.25826) $\\  
\hline
    \end{tabular}%
\end{table*}%

\section{Thermodynamic model based on a generalized cubic equation of state} 
\label{s:ThermoMod}

We present a thermodynamic model based on the generalized cubic EoS. First, we present the EoS and describe in detail how it is solved. Then, the correlations to evaluate the thermodynamic properties are presented and, finally, the Chung correlations for the transport properties are briefly described. 

\subsection{Generalized cubic equation of state}
\label{ss:eos}
We here consider the generalized formulation of a cubic EoS suggested by \citet{Cismondi} and already earlier by Mollerup~\citep{michelsen2004thermodynamic}
  \begin{equation} 
      p({v},T)=\frac{{R}T}{{v}-b}-\frac{a \alpha}{({v}+\delta_1b)({v}+\delta_2b)}\,.
      \label{eq:genEq}
  \end{equation}
The pressure $p$ is a function of the molar volume ${v}$ and the temperature $T$. ${R}$ denotes the universal gas constant with ${R} = 8314.472\;\si{J/(kmol\,K)}$. $a$ and $b$ represent the two traditional EoS parameters, considering attractive forces with $a$ and repulsive forces by the the effective molecular volume $b$. Both are determined by a proportionality factor and the critical properties $p_c$ and $T_c$ of the fluid, see \cref{tab:KimTabelle}. Further, $a$ is multiplied by a correction factor $\alpha$ that is a function of reduced temperature $T/T_c$ and the acentric factor $\omega$. It is worth noting that for $a=0$ and $b=0$, the cubic EoS collapses to the ideal gas law. As a consequence, mathematically, and also physically, the molar volume ${v}$ has to be larger than the co-volume $b$ (${v} > b$). The common cubic EoS can be described with special pairs of the values $\delta_1$ and $\delta_2$ , where $\delta_2$ is a supplementary parameter defined as $(1-\delta_1)/(1+\delta_1)$. Multiplying the denominator out results in the well-known and often used formulation of 
  \begin{equation} 
      p({v},T)=\frac{{R}T}{{v}-b}-\frac{a \alpha}{{v}^2 + u b {v} + w b^2 },
      \label{eq:genEq_old}
  \end{equation}
where $u=\delta_1+\delta_2$ and $w=\delta_1\delta_2$. However, the first formulation (\cref{eq:genEq}) yields simpler expressions of the derivations required for evaluating the thermodynamic properties (see \cref{ss:caloric}) than \cref{eq:genEq_old}. For the widely used EoS SRK and PR, the proportionality factor in $a$ and $b$ is constant and $\delta_1$, or respectively  $u$ and $w$, are constants with $\delta_1 = 1$ ($u= 1$, $w = 0$) for SRK and $\delta_1 = 1 + \sqrt{2}$ ($u=2$, $w=-1$) for PR. Hence, for SRK and PR a universal critical compressibility has been assumed, which is about 0.285 for SRK and 0.263 for PR~\citep{Kim}. Therefore, these two EoS are only well suited for a certain set of fluids with a corresponding similar critical compressibility. To overcome this limitation, \citet{Cismondi} suggested to evaluate the EoS parameters as a function of the critical compressibility resulting in the RKPR EoS. For the detailed evaluation of the EoS parameters see \cref{tab:KimTabelle}. For the RKPR, a different correlation than for SRK and PR is used to evaluate $\alpha$. Consequently, also the derivatives by temperature $\partial \alpha/\partial T$ and $\partial^2 \alpha/\partial T^2$, required for the evaluation of the thermodynamic properties, differ for the EoS. Concluding, all three cubic EoS SRK, PR, and RKPR can be described by \cref{eq:genEq}, where only the EoS parameters $a$, $b$, and $\delta_1$ as well as the evaluation of $\alpha$ changes. 

In order to solve the cubic EoS (\cref{eq:genEq}), the equation is reformulated using the dimensionless compressibility factor
  \begin{equation} 
      Z = \frac{p {v}}{{R}T}, 
      \label{eq:Z}
  \end{equation}
as well as  dimensionless expressions for $a$ and $b$ with $A=pa\alpha(RT)^{-2}$ and $B=pb(RT)^{-1}$. Therewith, one obtains from \cref{eq:genEq}
  \begin{equation} 
     Z =\frac{1}{1-B/Z}-\frac{A}{B}\;\frac{B/Z}{(1 + \delta_1B/Z)(1+ \delta_2B/Z)}\,.
      \label{eq:Zeq2}
  \end{equation}
Recasting results in the cubic form of all considered EoS in terms of $Z$, which reads
  \begin{equation} 
      Z^3 + a_2 Z^2+ a_1 Z + a_0 =0
      \label{eq:cubicFormZ}
  \end{equation}
with the coefficients
  \begin{equation} 
    a_2 = B(\delta_1+\delta_2-1)-1,
    \label{eq:a2}
  \end{equation}
  \begin{equation} 
      a_1 = A +\delta_1\delta_2 B^2-(\delta_1+\delta_2) B (B+1),
      \label{eq:a1}
  \end{equation}
    \begin{equation} 
      a_0 = -B(\delta_1\delta_2 B^2+\delta_1\delta_2 B+A). 
      \label{eq:a0}
  \end{equation}
The obtained cubic equation can be solved for real roots, which is well described in the  literature~\citep{elliott2012introductory,education_psu,Matheis}. A cubic equation has either one real and two imaginary roots or three real roots. At supercritical conditions usually only one real root is present. For this reason, we recommend first checking for the existence of one real root when evaluations are focused on supercritical conditions. Three real roots are generally associated with the two-phase region present at sub-critical conditions. As mentioned above, the volume has to be larger than the co-volume $b$ (${v} > b$), or expressed in terms of the compressibility factor $Z>B$. If the physical constraint $Z>B$ is full-filled, the smallest root represents the liquid state and the largest root the vapor or gas state. From a thermodynamic perspective, the center root is not stable and therefore physically meaningless. The correct root of the two physically stable roots can be identified by comparing the Gibbs energy, see \cref{ss:caloric}. An alternative approach is to take always the largest root, which usually corresponds to the vapor/gaseous state (exception see below). 

At very high and low pressures, the smallest root can be smaller than the co-volume ($Z\leq B$). In this case, there is only one physical meaningful root, which is the largest one. At high pressures, this root then corresponds to a liquid state, while at very low pressures it corresponds to the gaseous/vapor state. It is important to note that this can also occur in a clearly supercritical regime, such as for n-dodecane at the ECN-Spray A condition with $p=8\,\si{MPa}$ in the temperature range $T = 1097 - 1500 \, \si{K}$ using PR or RKPR. Consequently, a missing check for $Z> B$ results in high deviations for both EoS, while including the check yields moderate deviations of about 2.5\% for the RKPR, see \cref{fig:val_of}. 

The presented EoS can be extended to model a homogeneous mixture for an arbitrary number of components. To this end, the EoS parameter $a\alpha$ and $b$ have to be evaluated as a function of the mixture. Details can be found in the literature~\citep{Kim, Matheis, Matheis:2017dx, fathi2022large}. 

\subsection{Thermodynamic properties}
\label{ss:caloric}

Besides the correlation of density, pressure and temperature, also expressions  for thermodynamic properties, such as the internal energy ${e}$, entropy $s$, enthalpy $h$ and specific heats ${c}_p$ and ${c}_v$, are needed for CFD simulations. The evaluation of these quantities includes several thermodynamic derivatives, which can be solved using the departure function formalism. For more detailed information we refer to \citet{Poling, elliott2012introductory}, and for details on the formulations for the RKPR to \citet{Kim,fathi2022large}. The presented formulations are obtained from \citet{Matheis} and recast to be valid for the more generic formulation. 

For the internal energy this can be written as
  \begin{equation} 
      {e}({v},T)={e}_0 (T)+ \int_\infty^{{v}} \left(T \frac{\partial p}{\partial T}\bigg \vert_{{v}}-p\right) \mathrm{d} {v},  
      \label{eq:innEnergy}
  \end{equation}
where the subscript $0$ refers to the ideal reference state. The solution of the integral reads
  \begin{equation} 
      {e}-{e}_0= \left(a\alpha - T \frac{\partial a\alpha}{\partial T}\right) \,K,  
      \label{eq:innEnergy2}
  \end{equation}
where the term $K$ contains the following expression 
\begin{center}
\begin{equation}
  K = \frac{1}{b(\delta_1 - \delta_2)} ln \left( \frac{{v} + \delta_1 b}{{v} + \delta_2 b}\right).
  \label{eq:K}
\end{equation}
\end{center}%
Consequently, the enthalpy ${h}$ is calculated with 
\begin{equation} 
    {h} - {h}_0 = {e} - {e}_0 + p{v} - {R}T\,.
\label{eq:enthalpy}
\end{equation}%
resulting in the expression
\begin{equation} 
\begin{split} 
{h} - {h}_0 = \left(a\alpha - T \frac{\partial a\alpha}{\partial T}\right)\,K + p{v}- {R}T.
\end{split}
\end{equation} 
The entropy ${s}$ is obtained with 
\begin{equation} 
  {s}({v},T)={s}_0(T)+ \int_\infty^{{v}}\left(\frac{\partial p}{\partial T}\bigg \vert_{{v}}-\frac{{R}}{{v}}\right)\mathrm{d}{v} +{R}\ln(Z), 
\label{eq:entropy}
  \end{equation}
resulting in 
\begin{equation} 
{s} - {s}_0 = - K T \frac{\partial a\alpha}{\partial T} +  {R}\ln\left(1-\frac{b}{{v}}\right)\,.  
\end{equation}
Finally, the Gibbs energy $g$ is calculated using \cref{eq:enthalpy} and \cref{eq:entropy}
\begin{equation}
    {g} - {g}_0 = {h} - {h}_0  - T({s} - {s}_0) 
\end{equation}
\begin{equation}
    {g} - {g}_0 = a\alpha K + p{v} - {R}T\left(1 + \ln\left(1-\frac{b}{{v}}\right)\right). 
\end{equation}
As mentioned above, the Gibbs energy can be used to determine the most stable root out of three real roots. If the smallest root is larger than $B$ ($min(Z) > B$), then the smallest root represents the liquid state ($Z_l = min (Z)$) and the largest one the vapor state ($Z_v = max(Z)$). The relative difference between the Gibbs energy of the two solutions can be evaluated with:
\begin{equation} 
\begin{split}
d{g} = \frac{{g}_v - {g}_l}{{R}T} = \frac{A}{B (\delta_1 - \delta_2)} ln\left( \frac{(Z_l + \delta_1 B)(Z_v + \delta_2 B)}{(Z_l + \delta_2 B)(Z_v + \delta_1 B)} \right) \\
 - (Z_l - Z_v) + ln\left( \frac{Z_l - B }{Z_v -B}\right).
\end{split}
\end{equation}
If $d{g}<0$ (${g}_v < {g}_l$), the vapor state is stable. Contrary, if $d{g}>0$ (${g}_v > {g}_l$), the liquid state is stable. 

The heat capacity at constant volume ${c}_v$ is calculated with 
\begin{equation}
  ({c}_{v} - {c}_{v0}) = -T\frac{\partial^{2}a\alpha}{\partial T^2}K\,, 
  \label{eq:cv}
\end{equation}
 where ${c}_{v0}$ is evaluated using ${c}_{v0} = {c}_{p0} - R$. ${c}_{p0}$, the heat capacity at constant pressure at ideal reference state, is determined with the 7-coefficient or the 9-coefficient NASA polynomials. For the corresponding data for the 7-coefficient polynomials, we refer to \citet{Goos_Burcat}, and for the 9-coefficient ones to \citet{mcbride2002nasa}. Special attention should be paid to the fact that the polynomials are adapted for certain temperature ranges and that for a smooth calculation over several temperature ranges an appropriate implementation has to be done. Then, the heat capacity at constant pressure ${c}_p$ can be evaluated using
\begin{equation}
  c_{p} = c_{v} - T \left(\frac{\partial p}{\partial T} \bigg \vert _{{v}} \right)^{2} \bigg / \frac{\partial p}{\partial {v}} \bigg \vert _{_{T}}, 
  \label{eq:cp}
\end{equation}
with
\begin{equation}
  \frac{\partial p}{\partial T} \bigg \vert _{{v}} =
   \frac{{R}}{{v} -b} 
   + \frac{\partial a \alpha}{\partial T}\frac{1}{{D}}
  \label{eq:dpdT}
\end{equation}
and 
\begin{equation}
  \frac{\partial p}{\partial {v}} \bigg \vert _{p} =
   - \frac{{R}T}{({v} -b)^2}
    + \frac{a\alpha ( 2 {v} + (\delta_1 + \delta_2) b)}{{D}^2},
  \label{eq:dpdT}
\end{equation}
with the denominator $D$
\begin{equation}
   D= ({v} + \delta_1 b)({v} + \delta_2 b) = {v}^2 + (\delta_1 + \delta_2) b {v} + \delta_1 \delta_2 b^2. 
  \label{eq:dpdT}
\end{equation}

  The speed of sound $c$ is calculated using
  \begin{equation}
    c = \sqrt{\frac{{c}_{p}}{{c}_{v}}\frac{\partial p}{\partial {v}} \bigg \vert _{_{T}} \frac{{v}^{2}}{M}}.
   \label{eq:speed_of_sound}
  \end{equation}

  \begin{figure*}[!tb]
  \centering
    \subfigure[]{\includegraphics[trim={50 0 250 0}, clip, width=0.49\linewidth]{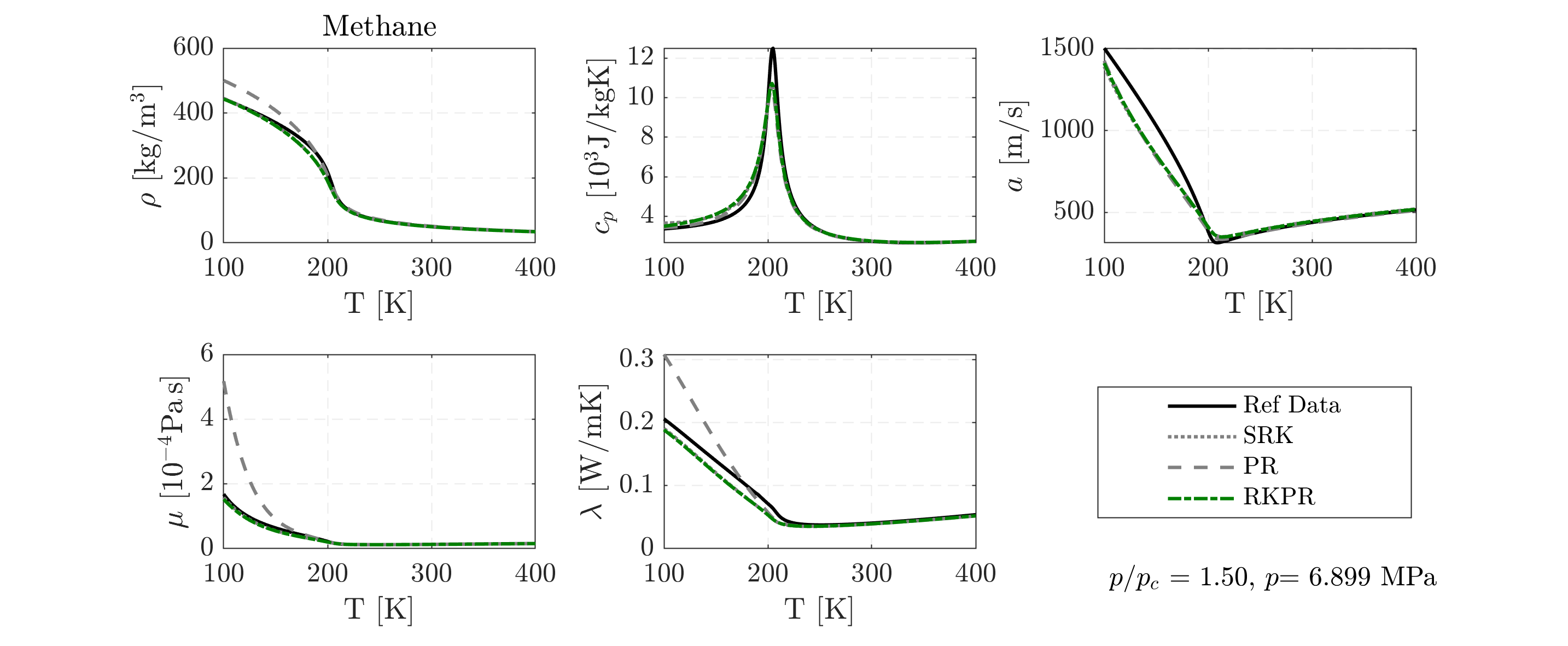}}
    \subfigure[]{\includegraphics[trim={50 0 250 0}, clip,width=0.49\linewidth]{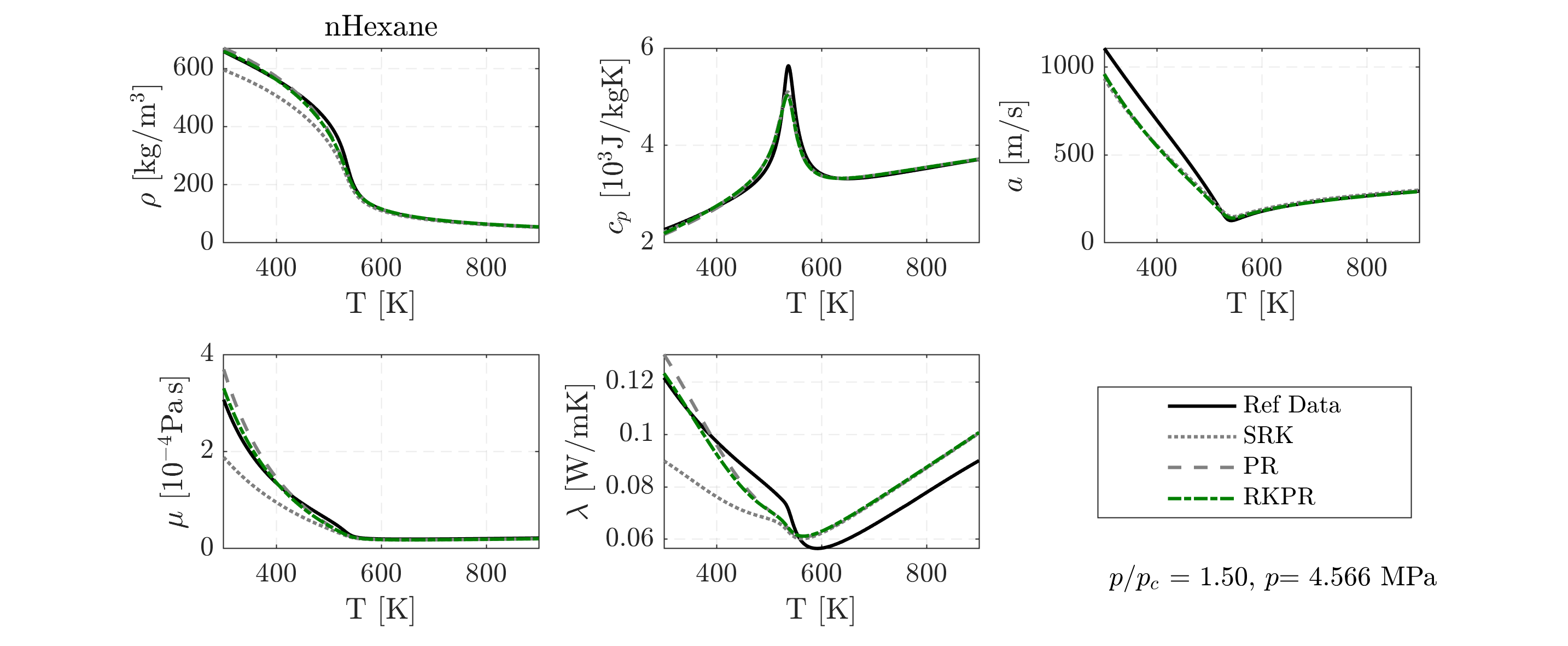}}\\
    \subfigure[]{\includegraphics[trim={50 0 250 0}, clip,width=0.49\linewidth]{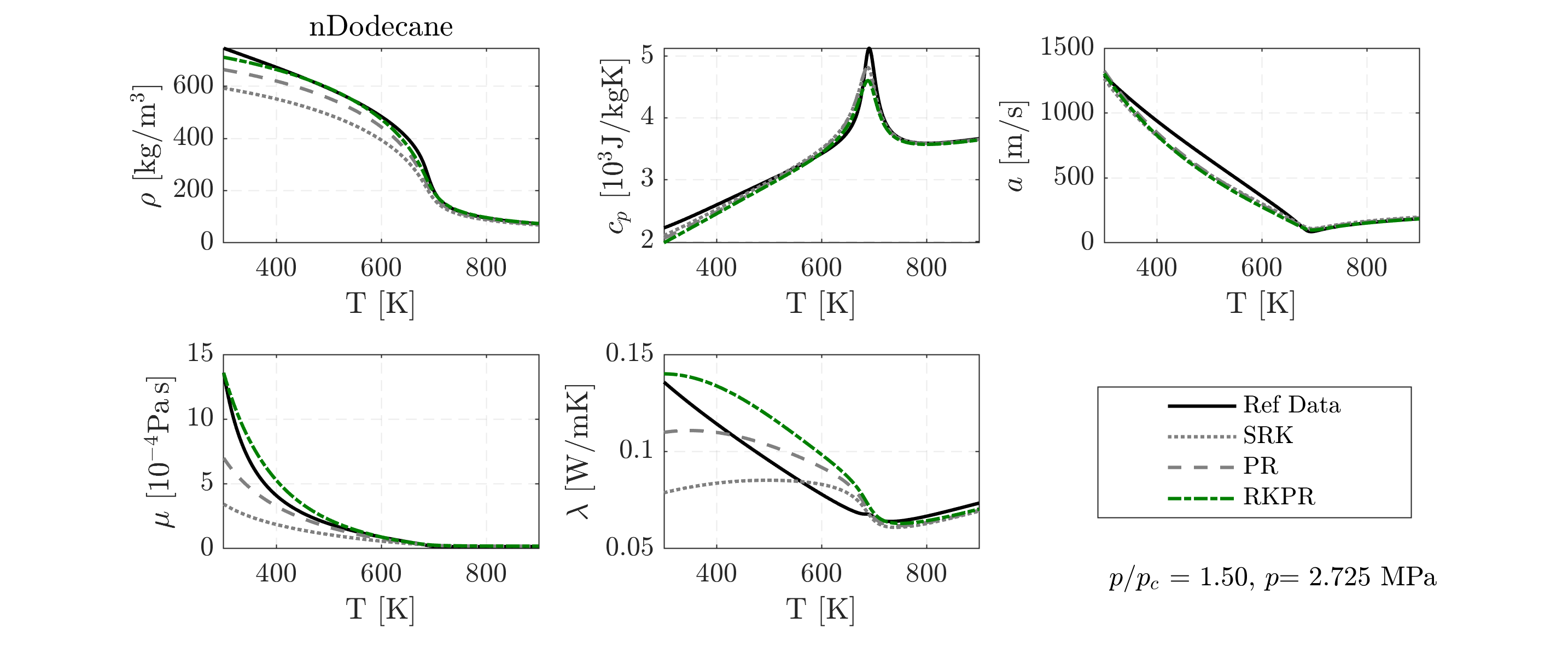}}
    \subfigure[]{\includegraphics[trim={50 0 250 0}, clip,width=0.49\linewidth]{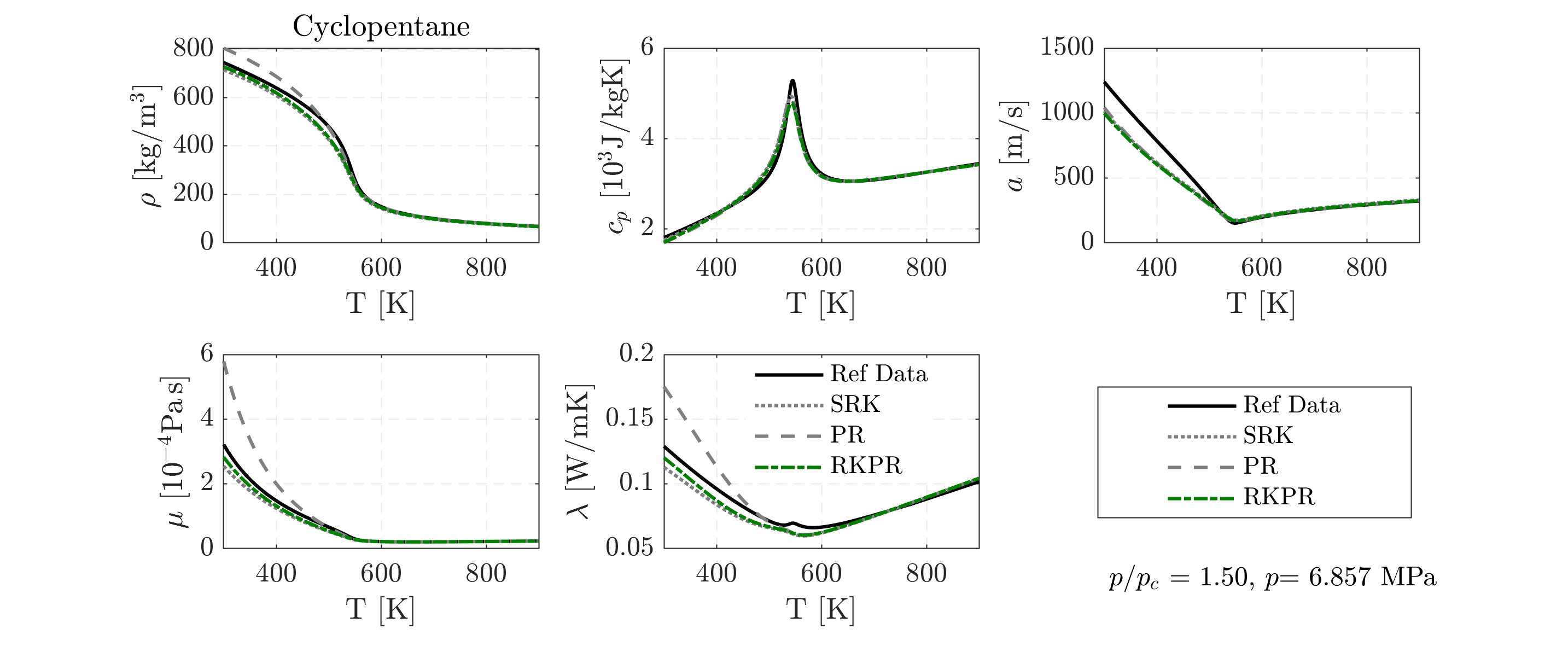}}
   \caption{
    Comparison of the modeled density $\rho$, heat capacity $c_p$, viscosity $\mu$ and heat conductivity $\lambda$ using the thermodynamic model employing different EoS with reference values from CoolProp~\citep{bell2014pure}. (a) Methane ($Z_c = 0.2863$), (b) n-hexane ($Z_c = 0.2664$), (c) n-dodecane ($Z_c = 0.2497$), and (d) cyclopentane ($Z_c = 0.2813$). All data has been evaluated at a relative pressure of $p/p_c = 1.5$. 
    }
   \label{fig:compare}
  \end{figure*}

\subsection{Transport properties viscosity and heat conductivity with the Chung correlations}
\label{ss:transport}

  For CFD simulations also suitable relations for the transport properties viscosity $\mu$ and heat conductivity $\lambda$ are necessary, where the correlations by \citet{Chung} are often employed~\citep{matheis2016volume,Matheis:2017dx, Matheis, fathi2022large,traxinger2018experimental, muller2016large}. Both quantities are composed of two terms referring to different pressure levels:
  \begin{equation}
    \mu = \mu_{k} + \mu_{p} \; \mathrm{and}
    \label{eq:mu}
  \end{equation}
  \begin{equation}
    \lambda = \lambda_{k} + \lambda_{p}.
    \label{eq:lambda}
  \end{equation}
  The first summand $\mu_{k}$, and $\lambda_{k}$ respectively, dominates at low pressures and is based on the Chapman-Enskog theory for diluted gases. The second term, $\mu_{p} $, and $\lambda_{p}$, dominates at higher pressures and is based on empirical correlations. The input for the model is composed of the fluid properties, the temperature, the density and the heat capacity ${c}_v$, where the latter only affects the evaluation of $\lambda$. For a detailed description see \citet{Chung,Poling}. 

\section{Assessment of the accuracy of the thermodynamic model}
\label{s:Accuracy}

  \Cref{fig:compare} shows the density $\rho$, the heat capacity $c_p$, the viscosity $\mu$  and heat conductivity $\lambda$ for the n-alkanes methane ($Z_c = 0.2863$), n-hexane ($Z_c = 0.2664$), n-dodecane ($Z_c = 0.2497$) and the cycloalkane cyclopentane ($Z_c = 0.2813$). All data has been evaluated at a relative pressure of $p/p_c = 1.5$. Overall, the thermodynamic modeling is able to reproduce the non-linear behavior for all depicted quantities. As expected, the fluids with a critical compressibility close to 0.285 are well described by SRK, while for n-hexane with $Z_c = 0.2664$ PR (optimized for $Z_c = 0.263$) gives good results. For n-dodecane RKPR yields the best results. In all cases, the density is modeled very well with the RKPR EoS. It is either comparable to SRK and PR or much better, if the critical compressibility differs from the values for which the two were designed. 

  The specific heat capacity $c_p$ is evaluated with \cref{eq:cp}. The peak at the pseudo-boiling is well captured by all EoS, but the maximum value is underestimated. 

  The overall behavior of the transport properties ($\mu$, $\lambda$) is well described by the Chung correlations. In comparison with the density evolution for the different cubic EoS, one can see that the error in the modeling of the density $\rho$ corresponds to the error in the modeling of these two quantities, see PR in \cref{fig:compare}~(a) or \cref{fig:compare}~(d). This is due to the fact that the density $\rho$ is an input parameter for the Chung model, which directly affects the calculation of the high pressure empirical terms $\mu_p$ and $\lambda_p$. Hence, the error of the Chung correlations increases with an increasing modeling error of the density. Further, for $\lambda$, the error of ${c}_v$ affects the calculation of $\lambda_k$.  As a consequence, the Chung correlations yield better results the more accurate $\rho$ and ${c}_v$ are modeled. 





\section{Supplementary material}
\label{s:SupMat}

   We provide an open source Python tool called \texttt{realtpl} (real gas thermodynamic python library) for the presented thermodynamic model. Additionally, we also provide the implementation of the generalized form in OpenFOAM.

\subsection{Python framework \texttt{realtpl}}

  Checking the accuracy of a thermodynamic model in advance is a central step before conducting CFD simulations. For different fluids as well as different  pressure and temperature ranges, such an evaluation can be complicated and especially time consuming. To this end, we have written an open source Python tool to easily compare the results obtained with the here described thermodynamic model based on cubic EoS. The Python tool is called \texttt{realtpl} standing for our real gas thermodynamic python library. It is directly coupled to the open source library CoolProp~\citep{bell2014pure} obtaining experimental reference data, as well as fluid properties, such as for example molar mass and critical properties. In addition, a database was created for the NASA coefficients, which is also directly coupled to \texttt{realtpl}. Using \texttt{realtpl}, thermodynamic modeling based on the cubic EoS PR, SRK, RKPR can be compared and also contrasted with the reference data from CoolProp. The current implementation is designed to evaluate results over a temperature range (with defined number of temperature steps) for a given pressure level. The data is displayed graphically and can also be exported to a \textit{csv} file for further processing. Moreover, also evaluations over temperature and pressure ranges can be done, which allows for table generation~\citep{Doehring,jafari2022exploring,Koukouvinis}. To this end, the ranges and also the step width can be specified as configuration parameter. Apart from that, this open source Python tool can serve as an inspiration for implementing the present model into an internal flow solver. 
  
  \Cref{tab:Performance} lists the process time for the evaluation of a representative configuration to provide estimates for the evaluation times using \texttt{realtpl}. The data has been evaluated for $10$, $10^2$, $10^3$ and $10^4$ temperature steps using a standard laptop (Intel i5, 7th generation).  \textit{Start-up} refers to the reading of the config files and corresponding fluid properties. \textit{Ref. data} stands for extracting the reference data from CoolProp. In the current version, the five quantities density, heat capacity, speed of sound, viscosity and heat conductivity are extracted. This extraction from CoolProp can not be vectorized and, thus, it has to be looped over the temperature steps. Therefore, the evaluation time required scales roughly linearly with the number of temperature evaluations. This is the main time consumer when about $\num{3e3}$ temperature evaluations are exceeded. For the thermodynamic model, we have here listed the average of all three cubic EoS named \textit{Thermo. model}. To improve performance, the implementation of the thermodynamic models has been recast as vectors, avoiding time-consuming loops. For this reason, there is no linear scaling of the evaluation process. Also at $10^4$ temperature evaluations the time per thermodynamic model is still about 0.05 s. Here it has to be noted that five quantities are evaluated. Among different cubic EoS, we have seen that it varies depending on how often the check for B and the Gibbs evaluation has to be done. The next contributions are then output and postprocessing related. \textit{Figures} refers to the visualization of all five quantities including the write out of the figures. For less than $\num{3e3}$, this is the most time consuming part. The last part \textit{Data-output} is the write out of all evaluated data to a \textit{csv} file and does not consume significant time. For the entire evaluation at e.g. $10^3$ temperature steps, approx. 6 s are required with writing out of the figures and approx. 2.2 s without.

  \texttt{realtpl} is available as a PIP Python package and also on github github.com/ttrummler/realtpl. 

  \begin{table}    
    \center
    \caption{Overall process time for different numbers of temperature evaluations (t-ev), for details see text. }
    \label{tab:Performance}
    \begin{tabular}{lrrrr} 
      \hline
      \multicolumn{1}{l}{} & 
      \multicolumn{1}{c}{10 t-ev [s]} & 
      \multicolumn{1}{c}{$10^2$ t-ev [s]} & 
      \multicolumn{1}{c}{$10^3$ t-ev [s]}& 
      \multicolumn{1}{c}{$10^4$ t-ev [s]} \\
      \hline
      Start-up & 0.020 & 0.020 & 0.020 & 0.020 \\
      Ref. data & 0.037& 0.1787& 1.489 &15.894\\  
      Thermo. model & 0.002 &0.002 &0.006 &0.046\\  
      Figures & 3.860 & 3.860 &3.860 &3.860\\ 
      Data-output & 0.018& 0.034 & 0.146 &0.365\\       
      \hline
      Total & 3.941 & 4.099 & 5.533 & 20.276\\
    \end{tabular}
  \end{table}

\subsection{Generalized Formulation in OpenFOAM$^{\tiny{\textregistered}}$}

  OpenFOAM is a widely used open source software for simulations, where currently the most recent versions are the foundation version OpenFOAM-10~\citep{OF10} and the ESI version OpenFOAM2206~\citep{OF2206}. In both, the PR EoS is available as \textit{PengRobinsonGas}. In some in-house extensions of OpenFOAM~\citep{traxinger2020pressure,traxinger2019single,Traxinger}, the SRK has been additionally implemented. Despite the identical structure of SRK and PR, these two EoS are often hard coded and thus, lead to code duplicates. Following the generalized formulation proposed above, we propose a more general implementation of cubic EoS to avoid code duplication and to improve readability. We provide this extension for OpenFOAM under github.com/ttrummler/realFOAM. In order to keep the traditional OpenFOAM code structure and to not change the input files, we have created three separate folders for the different EoS. 

\begin{figure}[!htb]
  \centering
    \subfigure{\includegraphics[trim={0 0 0 0}, clip, width=0.99\linewidth]{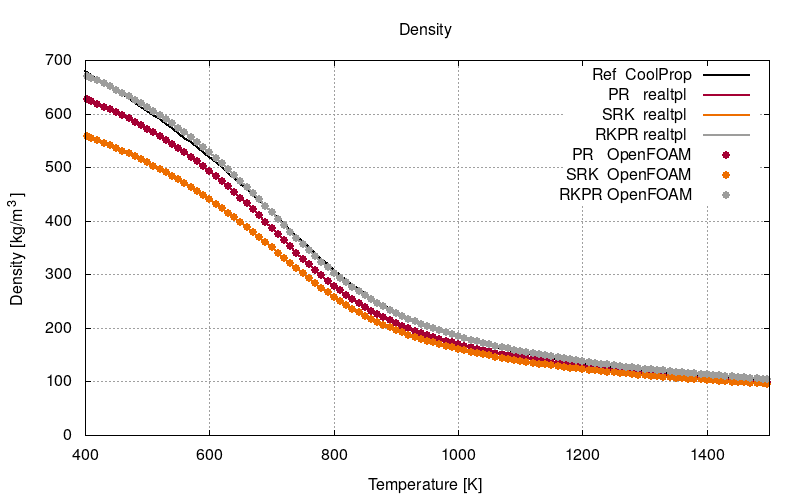}}
    \subfigure{\includegraphics[trim={0 0 0 0}, clip, width=0.99\linewidth]{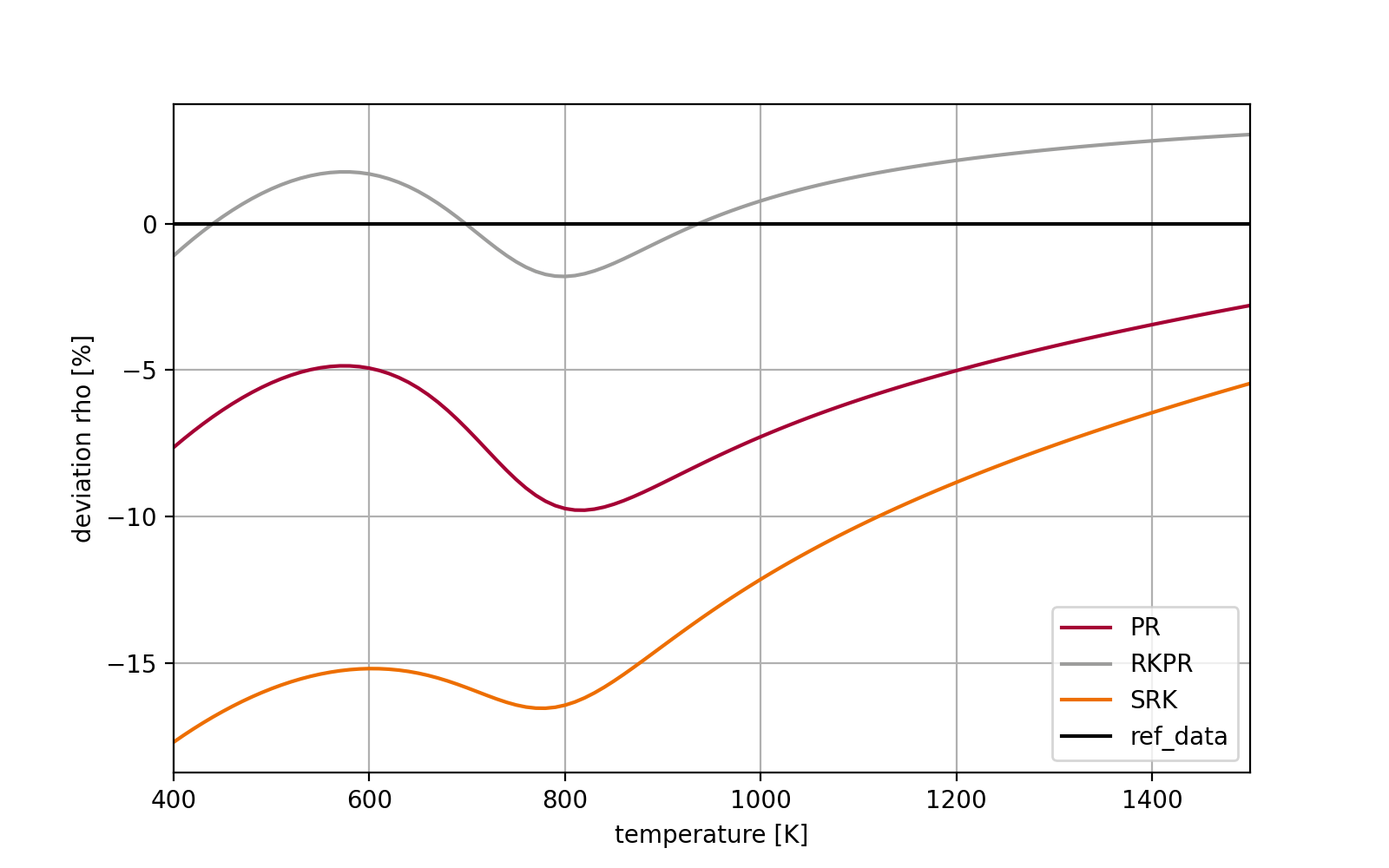}}
   \caption{
    Comparison of the modeled density for n-dodecane at a pressure of $p=8\,\si{MPa}$. Reference values are taken from CoolProp~\citep{bell2014pure}.
    }
   \label{fig:val_of}
  \end{figure}

  \Cref{fig:val_of} shows a validation of our OpenFOAM implementations comparing the density distribution with that obtained using the Python tool \texttt{realtpl}. As test configuration, we consider n-dodecane at a pressure of $p=8\,\si{MPa}$ and a temperature range of $T=500$ -- $1500 \,\si{K}$, matching the conditions of one operating point of the ECN Spray A~\citep{Matheis:2017dx, Koukouvinis}. For RKPR a very small deviation is visible, which is due to rounding errors. 

\section{Conclusions}
\label{s:Con}

  We have presented a thermodynamic model for real gas CFD simulations based on the generalized formulation of cubic EoS. Using this generalized formulation all of the well-known cubic EoS can be described with a particular value for $\delta_1$ ($\delta_2=(1-\delta_1)/(1+\delta_1)$). 
  We have provided a detailed presentation of the resulting generalized cubic equation in $Z$ and practical hints for solving it. To evaluate the thermodynamic properties, we presented formulations of the derivatives. The transport properties are modeled with the Chung correlations. The thermodynamic model allows for a modularized implementation of several EoS. 

  For the cubic EoS we have considered the well-known formulations SRK and PR. These two are specifically designed for an assumed critical compressibility factor and therefore their suitability is limited. Additionally, we also considered the RKPR, where the EoS parameters are functions of the critical compressibility factor. In this study, we have assessed the applicability of the three EoS for selected fluids and showed that the RKPR could be a good universally applicable choice for the EoS. In addition to this, we have demonstrated that overall the presented thermodynamic model can capture and reproduce the non-linear behavior of relevant thermodynamic quantities with an acceptable error. 

  As supplementary material for the paper we provide an open source Python tool that can be used to evaluate and compare the results for a wide range of different fluids.  Moreover, we also provide the implementation of the generalized EoS form in OpenFOAM. 


\section*{Acknowledgment}

  This project received funding by dtec.bw - Digitalization and Technology Research Center of the Bundeswehr - under the project MaST: Macro/Micro-simulation of Phase Decomposition in the Transcritical Regime.

\section*{Data Availability}

  The data that support the findings of this study are available from the corresponding author upon reasonable request. Further, the data supporting the findings of this study can be generated with the openly available open source code provided within this paper.

\section*{References}
\bibliography{pof}
\end{document}